\documentclass[%
 reprint,
superscriptaddress, 
 amsmath,amssymb,
 aps, prl, floatfix
]{revtex4-2}
\usepackage[utf8]{inputenc}
\usepackage[T1]{fontenc}
\usepackage{csquotes}
\usepackage{silence}
\usepackage{pgffor}
\newcommand{\vect}[1]{\boldsymbol{#1}}

\usepackage{xcolor} 
\usepackage[colorlinks=true,citecolor=blue,linkcolor=blue,urlcolor=blue]{hyperref}

\usepackage{pdfpages}
\usepackage{pgffor}

\newcommand{\chg}[1]{\textcolor{black}{#1}}

\makeatletter
\AtBeginDocument{\let\LS@rot\@undefined}
\makeatother

\begin{document}

\preprint{}


\title{Self-Organization and Cyclic Positioning of Active Condensates}
\author{Hossein Vahid}
\affiliation{Leibniz-Institut f{\"u}r Polymerforschung Dresden, Bereich Theorie der Polymere, 01069 Dresden, Germany}

\author{Jens-Uwe Sommer}
\email{jens-uwe.sommer@tu-dresden.de}
\affiliation{Leibniz-Institut f{\"u}r Polymerforschung Dresden, Bereich Theorie der Polymere, 01069 Dresden, Germany}
\affiliation{Technische Universit{\"a}t Dresden, Institut f{\"u}r Theoretische Physik, 01069 Dresden, Germany}

\author{Abhinav Sharma}
\email{abhinav.sharma@uni-a.de}
\affiliation{Faculty of Mathematics, Natural Sciences, and Materials Engineering: Institute of Physics, University of Augsburg, Universit{\"a}tsstraße 1, 86159 Augsburg, Germany}
\affiliation{Leibniz-Institut f{\"u}r Polymerforschung Dresden, Bereich Theorie der Polymere, 01069 Dresden, Germany}

\begin{abstract}
\chg{Cohesive active assemblies are often regulated by spatially heterogeneous nonequilibrium driving, such as gradients in motility, biochemical turnover, or mechanical activity.
Such heterogeneous driving can influence where condensates or cell collectives accumulate, how stable they are, and how they exchange material with their surroundings.
However, the minimal physical mechanisms by which activity gradients control the positioning and turnover of cohesive active matter remain unclear.
Here, we address this question using a model of attractive active Brownian particles (ABPs) in a spatially varying activity field.}
Using Brownian dynamics simulations, we show that these particles undergo liquid-gas phase separation, and spatially varying activity fields induce striking emergent dynamics. Attractive active droplets migrate up activity gradients, and at sufficiently high activity, they can fragment or evaporate into a dilute phase.
\chg{For finite clusters, evaporated ABPs can redistribute through the simulation box, reassemble into new clusters in lower-activity regions, and migrate again toward higher activity, giving rise to cyclic positioning through repeated nucleation, migration, evaporation, and reassembly.} These dynamics arise without biochemical feedback loops and rely only on the interplay of attraction, motility, and activity gradients.
\chg{Active--passive mixtures exhibit a core--shell-like organization, where the more cohesive ABPs condense into dense interiors and a peripheral layer dominated by passive particles. Furthermore, introducing the passive component shifts the steady-state localization of the condensates toward regions of higher activity compared to pure ABP systems.} 
\chg{Our findings identify a minimal physical route by which activity gradients can regulate the spatial localization, turnover, and cyclic repositioning of attractive active condensates.}
\end{abstract}

\date{\today}

\maketitle

\section{Introduction}

\chg{Living systems organize matter in space by combining cohesive interactions with spatially heterogeneous sources of nonequilibrium driving~\cite{karsenti2008, hyman2014, lin2015, scarpa2016, wedlich2018, weber2019, snead2019, lafontaine2021, ziepke2022, coupe2026}.
At the subcellular scale, cells assemble biomolecular condensates that compartmentalize reactions without membranes~\cite{brangwynne2011, feric2016, saha2016, jain2016, banani2017, snead2019, lafontaine2021, guilhas2020, linsenmeier2022, jambon2024, coupe2026}, and at the tissue scale, they coordinate the directed migration of multicellular groups during development, wound repair, and disease~\cite{ewald2008, carmona2008, mayor2010, theveneau2010, riahi2012, li2013, lin2015, scarpa2016, li2019, miskolci2021}. In both settings, structure is not controlled by equilibrium thermodynamics alone; cellular assemblies are continually remodeled by energy-consuming processes and by gradients in the surrounding chemical or mechanical environment. Understanding how such gradients bias the position, stability, and turnover of cohesive assemblies is therefore a general problem in nonequilibrium soft matter.}

\chg{Biomolecular condensates provide one important example of this principle. Many condensates are now understood as biomolecular assemblies with liquid-like properties, and their composition and material state are regulated by active biochemical processes~\cite{snead2019, lafontaine2021, linsenmeier2022, coupe2026}. ATP- and GTP-dependent enzymes, helicases, chaperones, phosphorylation cycles, and transport processes can all reshape condensate assembly and dissolution~\cite{brangwynne2011, feric2016, jain2016, patel2017, snead2019, guilhas2020, linsenmeier2022}. Moreover, ATP is not only an energy source but can also act directly as a hydrotrope that affects protein solubility at physiological concentrations~\cite{patel2017}.  These observations suggest that, in cells, spatial heterogeneity in metabolic or enzymatic activity can act as a spatially patterned nonequilibrium drive for condensate dynamics, even when there is no simple one-to-one correspondence between a condensate and a single control molecule.}

\chg{At larger scales, collective cell migration is likewise controlled by spatial gradients~\cite{carmona2008, theveneau2010, mayor2010, lin2015, scarpa2016, li2019, miskolci2021}. Migrating cell groups respond to chemokines, growth factors, extracellular-matrix composition, stiffness, confinement, and intercellular communication~\cite{giampieri2009, theveneau2010, lin2015, ellison2016, miskolci2021}. These cues regulate local protrusion, traction, polarity, and persistence, allowing multicellular assemblies to move more efficiently and more persistently than isolated cells. From a coarse-grained physical perspective, such guidance can often be viewed as a spatial modulation of the local magnitude of active forcing, that is, of how strongly individual units are driven out of equilibrium.}

\chg{The behavior of active particles in activity landscapes is already well understood for ideal and purely repulsive systems.
Spatial variations in swim speed generate density contrasts, accumulation in low-motility regions, and interfacial polarization even in the absence of explicit alignment interactions; abrupt motility steps also produce characteristic pressure imbalances~\cite{auschra2021, soker2021, row2020}.
What remains much less understood is how these ideas change once the particles are cohesive enough to condense.
In that regime, activity gradients no longer act on isolated particles only; they act on droplets or clusters whose interfacial polarization, internal turnover, and mechanical balance are all coupled to attraction.}

\chg{Here, we study this missing regime using attractive active Brownian particles (ABPs) in a spatially varying activity field. Here, the activity field could be interpreted generically as a gradient in nonequilibrium drive. Depending on context, it may represent spatially varying motility, traction, active stress, or the intensity of biochemical turnover that remodels an assembly. This formulation allows us to isolate the minimal physical consequences of combining cohesion with heterogeneous driving, without relying on the microscopic details of any one biological system.
We show that this minimal setting already produces a rich form of self-organization. Cohesive active particles form dynamic condensates that drift along activity gradients, can dissolve when driven strongly enough, and can subsequently reassemble elsewhere, leading to sustained turnover and spatial repositioning.
This behavior arises without any biochemical feedback or external control but as an emergent consequence of the interplay between attraction and activity gradients. Our results, therefore, identify a minimal physical route by which cohesive active matter can self-organize, relocate, and renew under heterogeneous nonequilibrium driving.
}

\section{Model}
We perform Brownian dynamics (BD) simulations of ABPs in a rectangular box of dimensions $L_x \times L_y \times L_z$.
Each ABP experiences the self-propulsion force $\vect{F}_{\rm a}^{i}=f_{\rm a}\vect{e}^{i}$, where $f_{\rm a}$ represents the magnitude of the active force, and $\vect{e}^{i}$ is a unit vector denoting the instantaneous propulsion direction of the particle. 
The equation of motion of the $i$-th ABP is expressed by 
$
\gamma_{\mathrm{t}} \, \dot{\vect{r}}^i = - \sum_j \nabla_{\vect{r}^i} U^{ij} + \vect{F}^i_{\mathrm{a}} + \vect{\xi}(t),
$
where \( \gamma_{\mathrm{t}} \) is the translational friction coefficient, \( U^{ij} \) the inter-particle interaction potential, and \( \vect{\xi}^i(t) \) a Gaussian white noise term satisfying \( \langle \vect{\xi}^i(t) \rangle = 0 \) and $\langle \xi^i_\alpha(t) \xi^j_\beta(t') \rangle = 2 \gamma_{\mathrm{t}}^{-1} k_{\mathrm{B}} T \delta^{ij} \delta_{\alpha \beta} \delta(t - t')$.
Here, $\alpha,\beta \in \{x,y,z\}$, and \( k_{\mathrm{B}} T=\beta^{-1} \) is the thermal energy, with $k_{\rm B}$ the Boltzmann constant and $T$ the temperature.
The orientation \( \vect{e}^i \) evolves via rotational diffusion given by $\dot{\vect{e}}^i = \vect{e}^i \times \vect{\eta}(t) $,
where \( \vect{\eta}(t) \) is a Gaussian white noise angular velocity vector having zero mean and correlations
$
\langle \eta^i_\alpha(t) \eta^j_\beta(t') \rangle = 2\gamma_{\mathrm{r}}^{-1} k_{\mathrm{B}} T \delta^{ij}\delta_{\alpha \beta} \delta(t - t'),
$
with \( \gamma_{\mathrm{r}} \) denoting the rotational friction coefficient.

Pairwise interactions of two ABPs at distance $r$ are modeled via the Wang–Frenkel potential~\cite{wang2020}:
$
U_{\rm WF}(r) = \epsilon \left[ \left( {\sigma}/{r} \right)^2 - 1 \right] \left[ \left( {r_{\rm c}}/{r} \right)^2 - 1 \right]^2 \Theta(r_{\rm c} - r),
$
where $r$ is their distance, $\epsilon$ the potential well depth, $\sigma$ the diameter of ABPs, $\Theta$ is the Heaviside step function, and $r_{\rm c}=2\sigma$ the cut-off distance.
The latter results in a Lennard-Jones-like behavior~\cite{wang2020}, and the potential quadratically vanishes at $r_{\rm c}$, ensuring the pair force disappears continuously.
We use reduced units by setting $\sigma=1$, $k_{\mathrm{B}}T=1$, and $\tau=\gamma_{\mathrm{r}}/k_{\mathrm{B}}T=1$, which define units of length, energy, and time, respectively.
The equations of motion are integrated using a timestep $5\times10^{-4}\tau$.
We set $\gamma_{\mathrm{r}}=\gamma_{\mathrm{t}}\sigma^2/3$, the P{\'e}clet number in homogeneous activity is defined as ${\rm Pe}=f_{\rm a}\sigma/k_{\mathrm{B}}T$, and $\rho_0$ is the bulk density of ABPs.
Simulations and visualizations are performed using \textsc{LAMMPS}~\cite{LAMMPS} and \textsc{Ovito}~\cite{stukowski2009} packages, respectively.

\begin{figure}[b!]
    \centering
    \includegraphics[width=1\linewidth]{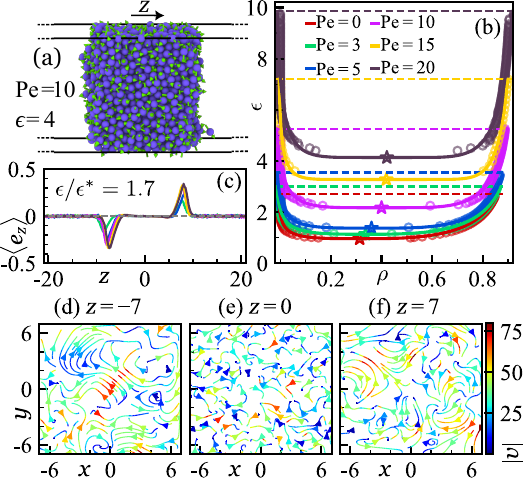}
    \caption{(a) A snapshot of the simulation box for ${\rm Pe}=10$ and $\epsilon=4$. The system exhibits phase coexistence, with a dense liquid-like slab at the center surrounded by a dilute vapor phase. (b) Phase diagrams as a function of attraction strength $\epsilon$. Circles indicate simulation points, solid lines are fits to Eq.~(S1) of SM, and dashed lines are the approximate $\epsilon$ values beyond which the system transitions to a densely-packed phase. Star symbols indicate the estimated critical points ($\rho^*$, $\epsilon^*$). (c) Profile of the average propulsion direction along $z$-axis, $\langle e_z(z) \rangle$, for various ${\rm Pe}$ at fixed $\epsilon/\epsilon^* = 1.7$. Legends of (b) also apply to (c). (d-f) Velocity streamlines at the interfaces of the condensed phase $z=\pm7$ and at the center of condensate $z=0$. Here, Pe$=10$ and $\epsilon=4$.}
    \label{fig:1}
\end{figure}


\section{Results}
To characterize the self-assembly and collective organization of attractive ABPs, we study their phase behavior by systematically varying $\epsilon$ across different Pe using direct coexistence simulations~\cite{prymidis2016, paliwal2017, omar2020, omori2024} (computational details are provided in the Supplemental Material (SM), Section SI). The phase diagrams in Fig.~\ref{fig:1}(b) reveal vapor–liquid coexistence, showing how cohesive interactions drive cluster formation. 
\chg{For a fixed Pe and a given attraction strength \(\epsilon\), the dilute and condensed coexistence densities are obtained from the two branches of the binodal.}
\chg{As \(\epsilon\) increases at fixed Pe, the horizontal separation between the dilute and dense branches of the binodal becomes larger, indicating an increasing density contrast between the coexisting phases.}
Increasing ${\rm Pe}$ broadens the coexistence region in both density and attraction strength, shifting the binodal to higher $\rho$ and higher $\epsilon$; in other words, stronger cohesion is needed to maintain phase separation at higher activity, and the critical density and attraction $(\rho^*, \epsilon^*)$ increase accordingly (see Table~SI of SM). These trends are consistent with experiments: motile \textit{E.\ coli} require stronger attractions to phase-separate than passive suspensions~\cite{schwarz2010,schwarz2012}, and higher ATP levels fluidize nucleoli, whereas ATP depletion stiffens them~\cite{brangwynne2011,feric2016}.

Beyond bulk densities, the structural organization at the vapor–liquid interfaces provides insight into self-assembly dynamics. We quantify particle alignment across the vapor-liquid interface through the orientation parameter $\langle e_z(z) \rangle = \frac{1}{N_z}\left\langle\sum_{i\in z}\vect{e}^i\cdot\hat{\vect{z}}\right\rangle_t$,
where \(N_z\) is the number of particles at position $z$, $\hat{\vect{z}}$ the unit vector along the $z$-axis, and \(\langle \cdot \rangle_t\) indicates the time average. 
Figure~\ref{fig:1}(c) shows that ABPs display pronounced outward polarization from the dense liquid phase at the interfaces. At fixed \(\epsilon/\epsilon^* = 1.7\), the alignment intensifies as ${\rm Pe}$ increases.
Alignment is negligible within bulk phases and along tangential directions \(\langle e_x(z) \rangle = \langle e_y(z) \rangle = 0\)~\cite{prymidis2016, paliwal2017, omar2020}.
\chg{
Instantaneous in-plane velocity streamlines, calculated from the \(x\)- and \(y\)-components of the particle velocities within \(xy\)-slices of thickness \(2\sigma\), are shown in Fig.~\ref{fig:1}(d--f).
The slices are centered at \(z=-7\), \(z=0\), and \(z=7\).
Near the interfaces at \(z=\pm7\), the in-plane motion is enhanced and contains locally circulating streamline patterns, whereas the central region at \(z=0\) shows slower and more disordered motion.
}

\chg{
To quantify mechanical stresses across the inhomogeneous system, we calculate the spatial profile of the normal pressure along the \(z\)-direction. The local normal pressure is decomposed into ideal, interaction, and active contributions as
\begin{equation}
P_{\rm tot}(z)
=
P_{\rm id}(z)
+
P_{\rm vir}(z)
+
P_{\rm act}(z).
\end{equation}
The ideal contribution is obtained from the local number density profile as
\begin{equation}
P_{\rm id}(z)=\rho(z)k_{\rm B}T .
\end{equation}
The interaction contribution \(P_{\rm vir}(z)\) is obtained from the slab-resolved configurational stress profile normal to the \(z\)-direction, using the same pressure sign convention as in the total normal pressure. For a slab of volume \(V_k=A\Delta z\), where \(A\) is the cross-sectional area and \(\Delta z\) is the bin width, the configurational contribution can be expressed in an Irving--Kirkwood form as~\cite{Ikeshoji01012003}
\begin{equation}
P_{\rm vir}(z_k)
=
\left\langle
\frac{1}{2V_k}
\sum_i\sum_{j\ne i}
\left[
- F^{ij}_z\,(z_b-z_a)
\right]
\right\rangle .
\end{equation}
Here, \(F^{ij}_z\) is the \(z\)-component of the force exerted by particle \(j\) on particle \(i\), and \(z_a\) and \(z_b\) denote the entry and exit positions of the Irving--Kirkwood contour through slab \(k\). Thus, each interacting pair contributes in proportion to the fraction of its contour contained within the slab.
}

\chg{
The active contribution is calculated from the slab-resolved polarization density~\cite{li2025surface}
\begin{equation}
m_z(z_k)=
\left\langle
\rho(z_k,t)\,\langle e_z\rangle_{k}(t)
\right\rangle_t .
\end{equation}
Here, \(\langle e_z\rangle_{k}(t)\) denotes the average orientation of particles located in the slab $k$ centered at \(z_k\) at time \(t\), and \(\langle\cdots\rangle_t\) denotes an average over the selected steady-state time window. For propulsion strength \(f_{\rm a}(z)\), this results in the local active body-force density
\begin{equation}
b_{\mathrm{act}}(z)
=
f_{\rm a}(z)\,m_z(z)
=
f_{\rm a}(z)
\left\langle
\rho(z,t)\,\langle e_z\rangle(z, t)
\right\rangle_t .
\end{equation}
The active contribution to the normal pressure is then obtained by integrating this body-force density along \(z\), $P_{\mathrm{act}}(z)
= -\int_{z_{\mathrm{ref}}}^{z}
b_{\mathrm{act}}(z')\,dz'$.
Here, \(z_{\mathrm{ref}}\) denotes the reference position used to set the additive constant of the integrated active contribution.
The integration is performed over the entire simulation domain, \(z_{\min}\le z\le z_{\max}\), with \(z_{\rm ref}=z_{\min}\).
The lower boundary \(z_{\min}\) is located in the dilute region, where \(b_{\mathrm{act}}(z_{\min})\simeq0\).
In practice, this integral is evaluated numerically using the trapezoidal rule on the same \(z\)-grid as the density and stress profiles.
}

\begin{figure}[t!]
    \centering
    \includegraphics[width=1\linewidth]{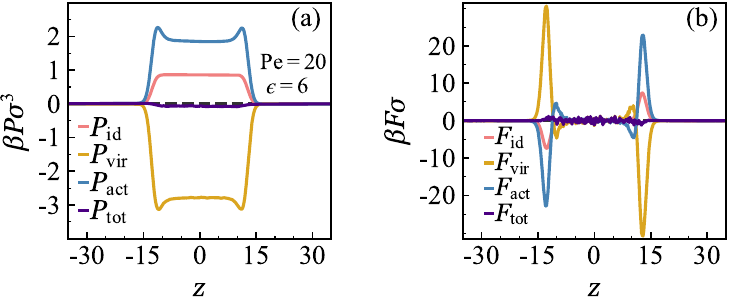}
    \caption{\chg{Pressure and force-balance decomposition for a representative system at \({\rm Pe}=20\) and \(\epsilon=6\). (a) Normal pressure profiles \(P_{\rm id}\), \(P_{\rm vir}\), \(P_{\rm act}\), and \(P_{\rm tot}\). The total pressure remains nearly flat because the ideal, virial, and active contributions compensate each other. (b) Corresponding slab-resolved force contributions \(F_{\rm id}\), \(F_{\rm vir}\), \(F_{\rm act}\), and \(F_{\rm tot}\), obtained using \(A\Delta z\) times the local force densities. The individual force contributions are localized at the interfaces and cancel, giving a zero net total force.
}}
    \label{fig:press}
\end{figure}


\chg{
We also compute the corresponding \(z\)-resolved force contribution in each slab. 
Thus, for the ideal and virial contributions,
\begin{subequations}
\begin{align}
F_{\rm id}(z)
&=
-A\Delta z
\frac{dP_{\rm id}}{dz}
=
-A\Delta z\,k_{\rm B}T
\frac{d\rho}{dz},
\label{eq:force_id}
\\
F_{\rm vir}(z)
&=
-A\Delta z
\frac{dP_{\rm vir}}{dz}.
\label{eq:force_vir}
\end{align}
\end{subequations}
For the active contribution, the slab-resolved force is obtained directly from the active body-force density,
\begin{equation}
F_{\rm act}(z)
=
A\Delta z\, b_{\rm act}(z).
\end{equation}
Here \(\Delta z=0.1\) is the bin width used in the profile calculation. The total slab-resolved force is then
\begin{equation}
F_{\rm tot}(z)
=
F_{\rm id}(z)+F_{\rm vir}(z)+F_{\rm act}(z).
\end{equation}
}

\chg{
Figure~\ref{fig:press} shows the decomposition of the normal pressure and the corresponding slab-resolved force contributions for a representative system with \(5000\) ABPs in a \(15\times15\times80\) simulation box at \({\rm Pe}=20\) and \(\epsilon=6\).
Panel~(a) shows the ideal, virial, and active pressure profiles.
The ideal pressure increases within the dense phase, indicating a higher local density.
The virial pressure is negative and develops pronounced minima at the interfaces, because attractive interactions generate a tensile configurational stress where the density changes rapidly.
In contrast, the active pressure shows positive interfacial peaks. These peaks result from the interfacial polarization of ABPs, which produces a localized active body-force density \(b_{\rm act}(z)=f_{\rm a}(z)m_z(z)\).
The active pressure, obtained by integrating this body-force density, therefore varies most strongly across the interfacial regions.
Although the individual ideal, virial, and active pressure contributions are large, their sum remains nearly flat, indicating that the system is close to mechanical balance in the steady state.
}

\chg{The corresponding bin-resolved forces in Fig.~\ref{fig:press}(b) are localized at the interfaces because they are proportional to the gradients of the pressure profiles.
The large positive and negative peaks in \(F_{\rm id}\), \(F_{\rm vir}\), and \(F_{\rm act}\), therefore, reflect the sharp interfacial variations of the corresponding pressure terms.
Their strong cancellation gives a nearly vanishing total force \(F_{\rm tot}\), consistent with the nearly constant total pressure in Fig.~\ref{fig:press}(a).
}

\begin{figure}[t!]
    \centering
    \includegraphics[width=1\linewidth]{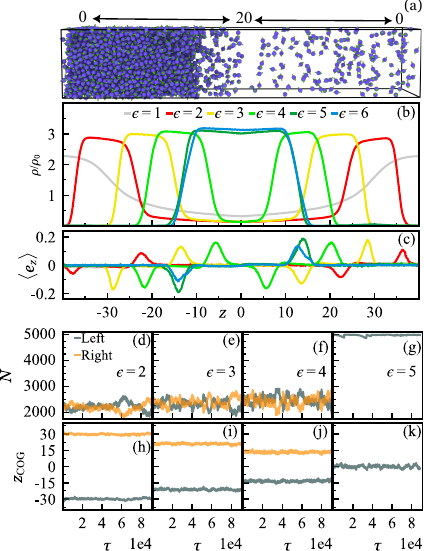}
    \caption{(a) Initial configuration at time $t=0$, showing a dense droplet with $\epsilon=4$ and $N=5000$ placed on the left side of the box ($z < -10$), where activity is lowest. The activity field is given by $f_{\rm a}(z) = 20(1 - |z|/40)$, creating a linear gradient symmetric about $z=0$.
    (b) Steady-state normalized density profiles $\rho(z)/\rho_0$ for varying $\epsilon$. \chg{Additional results for other $N$ and activity gradient steepness are presented in Fig.~S3 of SM.}
   (c) Steady-state profiles of the ABP propulsion direction projected along the $z$-axis, $\langle e_z(z) \rangle$.
   \chg{(d)--(g) Time evolution of the droplet size \(N(t)\), defined as the number of ABPs in each daughter droplet, for \(\epsilon=2,3,4,\) and \(5\), respectively.
    The two curves correspond to the left and right droplets.
    (h)--(k) Corresponding time evolution of the droplet center-of-geometry along the activity-gradient direction, \(z_{\rm COG}(t)\), for the same values of \(\epsilon\). For $\epsilon=5$, only a single trace is present because the droplet does not split.}}
    \label{fig:2}
\end{figure}

We now turn to how an attractive ABP condensate responds to activity gradients. To this end, we conduct BD simulations of a condensate of $5000$ ABPs placed at $-40\leq z\leq -10$ within a periodic box of size $15 \times 15 \times 80$, with activity varying only along the $z$-axis, $f_{\rm a}(z) = 20\left(1 - |z|/40\right)$, peaking at $z=0$ and vanishing at $z=\pm 40$ (Fig.~\ref{fig:2}(a)). Figure~\ref{fig:2}(b) shows steady-state density profiles along $z$ for various $\epsilon$. At weak attraction ($\epsilon=1$), activity overwhelms cohesion; the droplet evaporates, and ABPs disperse, accumulating in low-activity regions.
At strong cohesion ($\epsilon>5$), the parent droplet remains intact and migrates as a single self-propelled entity to $z=0$ (highest-activity region).

Interestingly, at moderate attraction strengths ($\epsilon=2-4$), droplets climb the gradient but are fragile; the parent droplet fissions into two daughter droplets that settle in intermediate-activity regions where active and attractive forces roughly balance.
\chg{This macroscopic behavior is accompanied by continuous microscopic turnover: ABPs detach from and rejoin the condensates while the droplets remain statistically localized.
Such continuous assembly and disassembly is characteristic of \textit{living clusters}.} Active systems often form living clusters that continually assemble and break apart under the balance of self–propulsion and cohesion~\cite{ginot2018, mognetti2013, caprini2024, palacci2013, bechinger2016}, as seen in active colloids and light–activated Janus living crystals~\cite{palacci2013}. 
Analogous turnover and reorganization processes occur in cell collectives when adhesion competes with motility~\cite{alert2020}. As shown in Figs.~\ref{fig:2}(c), ABPs at the condensate interfaces exhibit an outward orientation from the dense phase.

\chg{
Figures~\ref{fig:2}(d)--(k) quantify this behavior by tracking the droplet size $N(t)$ and its center-of-geometry $z_{\rm COG}(t)$ along the activity gradient over the final $10^5\tau$ of simulation trajectories. At intermediate attraction strengths $\epsilon=2$--$4$, the daughter droplets exhibit steady-state turnover. Their sizes fluctuate around a mean value, indicating continuous exchange of ABPs with the surrounding dilute phase, and their centers of geometry remain statistically stable.
Increasing the attraction strength to $\epsilon \ge 5$ stabilizes the 
condensates against activity-induced deformation and particle loss. The corresponding time traces reveal larger condensed masses as well as 
substantially smaller fluctuations in both $N(t)$ and $z_{\text{COG}}(t)$.
}

\begin{figure}[t!]
    \centering
    \includegraphics[width=1\linewidth]{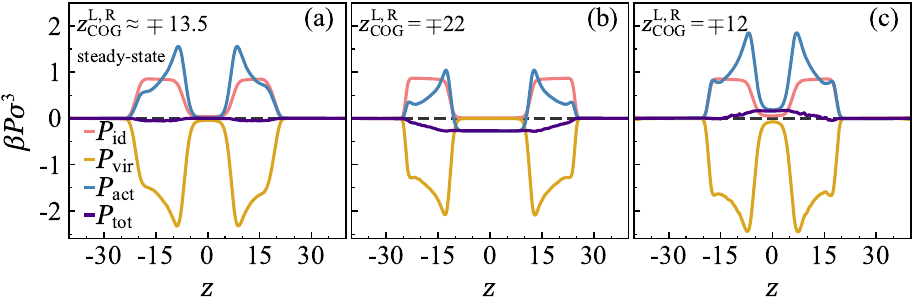}
\caption{\chg{
Pressure decomposition in the activity gradient \(f_{\rm a}(z)=20(1-|z|/40)\) at \(\epsilon=4\).
(a) Steady-state pressure profiles for unconstrained condensates.
(b) Pressure profiles for condensates constrained farther from the activity maximum, \(z_{\rm COG}^{\rm L,R}=\mp 22\).
(c) Pressure profiles for condensates constrained closer to the activity maximum than their steady-state position, \(z_{\rm COG}^{\rm L,R}=\mp 12\).
The curves show the ideal, virial, active, and total normal pressure contributions, \(P_{\rm id}\), \(P_{\rm vir}\), \(P_{\rm act}\), and \(P_{\rm tot}\), respectively.
}}
    \label{fig:press2}
\end{figure}

\chg{
Figure~\ref{fig:press2} shows the pressure decomposition for attractive ABP condensates at \(\epsilon=4\) in the spatial activity gradient
\(f_{\rm a}(z)=20\left(1-|z|/40\right)\).
Panel~(a) shows the steady-state configuration, where the condensates are not constrained.
Panels~(b) and (c) show constrained configurations in which the condensates are held away from this steady-state position by re-centering them at each time step.
In panel~(b), the condensates are kept farther from the activity maximum, \(z_{\rm COG}^{\rm L,R}=\mp22\), whereas in panel~(c), they are kept closer to the activity maximum, \(z_{\rm COG}^{\rm L,R}=\mp12\).
The individual pressure contributions are asymmetric across each daughter droplet because the inner and outer interfaces experience different local activities and interfacial structures.
Here, the inner and outer interface denote the side of each condensate facing the high- and low-activity regions, respectively.
The ideal pressure follows the density profile and is larger where the local density is higher.
Since the higher activity near the inner interfaces enhances particle exchange and particle loss from the condensate, the density, and therefore \(P_{\rm id}\), can decrease from the outer to the inner side.
The virial pressure is negative and develops pronounced interfacial minima due to attractive configurational stresses.
The active pressure shows interfacial peaks arising from the polarization-induced active body-force density.
}

\chg{
In the steady state [Fig.~\ref{fig:press2}(a)], these asymmetric contributions compensate so that
\(P_{\rm tot}=P_{\rm id}+P_{\rm vir}+P_{\rm act}\) is nearly flat, indicating mechanical balance.
When the condensates are displaced from this position, the compensation is incomplete and results in nonuniform \(P_{\rm tot}\).
For condensates held farther from the activity maximum [Fig.~\ref{fig:press2}(b)], the total-pressure imbalance corresponds to an inward mechanical drive, which would move the condensates toward the high-activity region if they were not re-centered.
In contrast, when the condensates are held closer to the activity maximum than their steady-state position [Fig.~\ref{fig:press2}(c)], the imbalance changes sign, indicating a restoring tendency away from the center.
}

\chg{
To quantify the mechanical imbalance, we determine the two interface positions of each condensate from the time-averaged density profile.
The net inward force on a condensate is calculated from the total-pressure difference between its outer and inner interfaces, $ F_{\rm net} = A\left[
P_{\rm tot}^{\rm out} - P_{\rm tot}^{\rm in} \right],$
where \(P_{\rm tot}^{\rm out}\) and \(P_{\rm tot}^{\rm in}\) are the total pressures at the outer and inner interfaces, respectively.
With this convention, \(F_{\rm net}>0\) corresponds to a force toward the activity maximum, whereas \(F_{\rm net}<0\) corresponds to a force away from it.
To this end, we calculate the net force 
from constrained simulations in which the two condensates are held at prescribed positions, \(z_{\rm COG}^{\rm L,R}=\mp d\), by re-centering them at each time step. 
The net force is obtained for each imposed distance \(d=|z_{\rm COG}|\), and we report the average net force by averaging over the left and right condensates.
The position-dependent drift velocity can also be calculated independently from unconstrained simulations. For each value of \(\epsilon\), we perform 400 simulations with independent random seeds and follow the motion for \(2\times 10^{4}\tau\). At each recorded frame, the centers of geometry of each daughter droplet, \(z_{\rm COG}^{\rm L}(t)\) and \(z_{\rm COG}^{\rm R}(t)\), are calculated.
The inward drift velocity is obtained from finite-time displacements,
\begin{equation}
  v_{\rm in}(d)
=
-\left\langle
\frac{d(t+\tau)-d(t)}{\tau}
\right\rangle_{d(t)\in[d-\Delta d/2,d+\Delta d/2]}, 
\end{equation}
where the average is taken over both condensates and all independent simulations. Since the condensates can exchange particles with the dilute phase and their sizes vary with time, the velocity average is weighted by the instantaneous condensate size, $ \langle v_{\rm in}(d)\rangle
= \frac{\sum_s N_{s}\,v_s(d)}{\sum_s N_{s}},$ where \(N_{s}\) is the number of particles in the condensate associated with sample \(s\). With this convention, positive \(v_{\rm in}\) denotes migration toward \(z=0\).
}

\begin{figure*}[t!]
    \centering
    \includegraphics[width=0.8\linewidth]{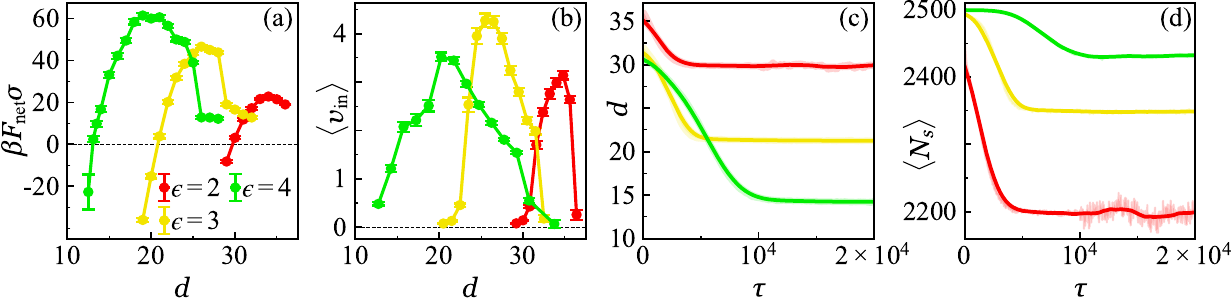}
    \caption{\chg{
Mechanical force and drift dynamics of attractive active condensates in the activity gradient of \(f_{\rm a}(z)=20(1-|z|/40)\). 
(a) Net inward force \(F_{\rm net}\) as a function of the imposed condensate distance from the activity maximum, \(d=|z_{\rm COG}|\), for different attraction strengths \(\epsilon\).
Positive values correspond to a force toward the activity maximum at \(z=0\).
(b) Position-dependent inward drift velocity \(\langle v_{\rm in}(d) \rangle\), plotted in units of \(10^{-3}\sigma/\tau\), calculated from unconstrained migration simulations.
For each \(\epsilon\), \(400\) independent simulations are run for \(2\times10^{4}\tau\).
The velocity average is weighted by the instantaneous condensate size \(N_s\).
(c) Time evolution of the mean condensate distance $d$ from the activity maximum, showing the migration toward the activity maximum and the approach to a steady-state position.
(d) Corresponding time evolution of the mean condensate size, \(\langle N_{s}\rangle\), showing that the condensates remain finite while migrating and approaching their steady-state positions. Pale-shaded regions indicate the standard error across independent simulations.}}
    \label{fig:fore_vel}
\end{figure*}

\chg{
Figure~\ref{fig:fore_vel} shows that the mechanically calculated forces and the independently calculated drift velocities are consistent. In the range where the condensates migrate toward the activity maximum, \(F_{\rm net}\) is positive and \(v_{\rm in}\) is also positive. Both quantities exhibit a maximum at intermediate distances and decrease as the condensates approach their steady-state positions. The two curves are not expected to collapse point by point because the drift velocity also depends on an effective mobility, which varies with condensate size, shape, and particle exchange with the dilute phase.
The time-dependent trajectories in Fig.~\ref{fig:fore_vel}(c) provide the corresponding time view of this migration process.
Starting from condensates located away from the activity maximum, the mean distance \(d(t)\) decreases with time, demonstrating net inward motion toward the high-activity region.
At later times, \(d(t)\) approaches a plateau, indicating that the condensates reach statistically steady positions.
The plateau position depends on \(\epsilon\), consistent with the force and velocity profiles in Fig.~\ref{fig:fore_vel}(a) and (b): condensates migrate inward over the range where \(F_{\rm net}>0\) and \(v_{\rm in}>0\), and slow down as they approach their steady-state locations.
Figure~\ref{fig:fore_vel}(d) shows the corresponding evolution of the mean condensate size, \(\langle N_{s}\rangle\).
As the condensates migrate toward the activity maximum, \(\langle N_{s}\rangle\) decreases, indicating partial loss of condensed particles due to stronger activity in the central region.
At later times, when the condensates approach their steady-state positions, \(\langle N_{s}\rangle\) also reaches an approximately constant value.
For the weakest attraction shown, \(\epsilon=2\), the fluctuations around the mean are larger than for higher \(\epsilon\), indicating weaker cohesion and stronger exchange of ABPs between the condensates and the surrounding dilute phase.}



\begin{figure}[b!]
    \centering
    \includegraphics[width=1\linewidth]{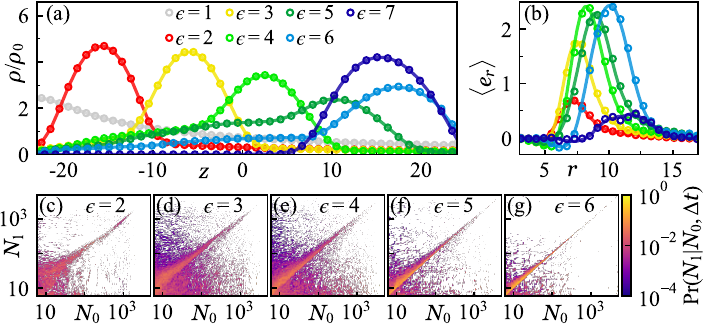}
    \caption{(a) Steady-state density profile of ABPs along the z axis. The activity field is given by $f_{\rm a}(z) = f_{\rm a}^*(z + 25)/50$, with $f_{\rm a}^*=20$. The system along the z-axis is non-periodic. (b) $\langle e_{r}\rangle=\langle\Sigma_{i\in r} \vect{e^i}\cdot\vect{\hat{r}}\rangle_t$ for the largest cluster as a function of $r$ from the cluster center-of-geometry, averaged over time. (c)-(g) Transition matrix $P(N_1 \vert N_0, \Delta t)$ for varying $\epsilon$, with $\Delta t=10\tau$. Results for $f_{\rm a}^*\in\{10,40,80\}$ are presented in Figs.~S5 and S6 of SM.
    }
    \label{fig:4}
\end{figure}

To further investigate this phenomenon for finite clusters, we conduct simulations of \(5000\) attractive ABPs under a spatially varying activity field.
The ABPs are randomly initialized in a cubic box of size $50^3$, with harmonic, repulsive walls that keep them within $z\in[-25,25]$ (cf. Sec.~SIV of SM). Periodic boundary conditions are imposed along the $x$ and $y$ directions.
The propulsion force of each ABP increases linearly along the $z$-axis, defined by $f_{\rm a}(z) = 20(z + 25)/50$.
Figure~\ref{fig:4}(a) shows the resulting steady-state density profiles for different attraction strengths $\epsilon$.
At $\epsilon=1$, the attraction is weak to support stable aggregates, and ABPs mainly accumulate in the low-activity region.
At intermediate attraction strengths $\epsilon=2-6$, finite clusters nucleate preferentially in the low-activity region and migrate toward higher activity (see Movie~1 of the SM).
The position of the density peak shifts progressively toward higher \(z\) as \(\epsilon\) increases, showing that stronger cohesion allows clusters to penetrate farther into the high-activity region.
For sufficiently strong attraction, as in the \(\epsilon=7\) case, the system forms a densely packed aggregate that remains localized in the highest-activity region with strongly suppressed particle loss.

\begin{figure*}[t!]
    \centering
    \includegraphics[width=1\linewidth]{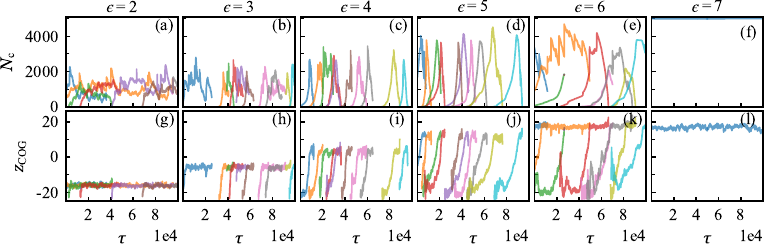}
\caption{\chg{
Cluster tracking in the activity gradient for the systems shown in Fig.~\ref{fig:4}, calculated during the final \(10^5\tau\).
(a)--(f) Time evolution of the cluster size \(N_{\rm c}(t)\) for tracked clusters at \(\epsilon=2,3,4,5,6,\) and \(7\), respectively.
(g)--(l) Corresponding center-of-geometry position \(z_{\rm COG}(t)\) along the activity-gradient direction.
Different colors correspond to different tracked clusters.
For intermediate attractions, \(\epsilon=3\)--\(6\), individual clusters can grow to large sizes, migrate to higher \(z\), and then shrink and disappear while new tracked clusters appear at later times.
Cluster trajectories that terminate within the plotted time window before \(N_{\rm c}\) decreases to zero indicate that the corresponding cluster has merged with another.
At the strongest attraction shown, \(\epsilon=7\), the system forms an almost fully condensed aggregate with little particle loss over the observation time.
Animations of the simulations are provided in Movie~1 of the SM.
}}
\label{fig:persistent_clusters}
\end{figure*}

The migration toward higher activity is associated with interfacial orientational polarization.
As shown in Fig.~\ref{fig:4}(b), ABPs at the droplet interfaces orient outward from the dense phase (Fig.~\ref{fig:4}(b)); see also Refs.~\cite{prymidis2016, paliwal2017, omar2020}. 
\chg{ABPs} on the high-activity side experience stronger propulsion, and the outward orientation ordering of interfacial ABPs generates a net force toward regions of higher activity.
Qualitatively, it is analogous to active colloidal dimers with opposite orientations, which tend to accumulate in high-activity regions~\cite{vuijk2022}.
At $\epsilon=7$, the condensates are \chg{densely packed} and the interfacial polarization decreases significantly compared to liquid-like condensates $\epsilon<7$ (see Fig.~\ref{fig:4}(b)). However, this weak polarization is still sufficient to drive the condensate to high-activity regions.
We quantify the aggregation--fragmentation dynamics over a short time interval of $10\tau$ using transition matrices,
\({\rm Pr}(N_1 \mid N_0,\Delta t=10\tau)\), shown in Figs.~\ref{fig:4}(c)--(g).
These matrices give the conditional probability that a cluster of size \(N_0\) evolves into a cluster of size \(N_1\) after a time interval \(\Delta t=10\tau\).
The probability is concentrated near the diagonal \(N_1\simeq N_0\), indicating that most clusters change size gradually over this time scale rather than undergoing abrupt dissolution and reformation.
The off-diagonal weight near the diagonal corresponds to the gain or loss of particles by exchange with the surrounding dilute phase. \chg{For weaker attractions, the probability distribution is broader, indicating stronger cluster-size fluctuations and more frequent particle exchange.
As \(\epsilon\) increases, the probability becomes more concentrated near the diagonal \(N_1\simeq N_0\), and probability weight appears at larger values of both \(N_0\) and \(N_1\), showing that stronger cohesion supports larger clusters.}

\chg{The finite-cluster dynamics for cases shown in Fig.~\ref{fig:4} should be distinguished from the planar-slab geometry discussed above, where the dense phase spans the simulation box in the two directions perpendicular to the activity gradient.
In that geometry, the two condensates are extended slabs located on opposite sides of the activity maximum.
When ABPs evaporate from the interfaces facing the high-activity region, they remain confined to the region between the two extended condensates and cannot bypass them to reach the low-activity sides.
Here, in contrast, the condensates are finite clusters in the simulation box.
Evaporated ABPs can move around the clusters and redistribute freely through the dilute phase.
Thus, this geometry provides a stricter test of whether localized condensates are maintained by continuous cluster turnover rather than by the geometric constraint of two extended slabs.}

\chg{Figure~\ref{fig:persistent_clusters} shows the time evolution of the cluster size \(N_{\rm c}(t)\) and the corresponding center-of-geometry position \(z_{\rm COG}(t)\) during the final \(10^5\tau\).
Different colors denote different tracked clusters, so that the size and position of the same cluster can be compared between the upper and lower panels.
For weak attraction, \(\epsilon=2\), the clusters remain relatively small and fluctuate near the lower-activity region.
For intermediate attractions, \(\epsilon=3\)--\(6\), clusters typically appear in the lower-activity region, grow there, and then migrate toward larger \(z\), corresponding to regions of higher activity.
The largest value of \(z_{\rm COG}\) reached by a cluster increases with \(\epsilon\), showing that stronger cohesion allows clusters to penetrate farther into the high-activity region before losing stability.
Comparing \(N_{\rm c}(t)\) with \(z_{\rm COG}(t)\) for each tracked cluster shows that cluster nucleation occurs mainly in the lower-activity region, whereas shrinkage and evaporation occur after the cluster has migrated into higher-activity regions. New clusters appear at later times in the lower-activity regions from ABPs that had previously evaporated and redistributed through the dilute phase.
Cluster trajectories that end within the plotted time window without \(N_{\rm c}\) approaching zero indicate merging with another cluster.
The dynamics, therefore, involve continuous cluster birth in low-activity regions, growth and migration toward higher activity, and subsequent evaporation or merging.
For the strongest attraction shown, \(\epsilon=7\), ABPs remain in an almost fully condensed aggregate with little particle loss, indicating that strong cohesion suppresses the evaporation observed at intermediate attraction strengths.
}

\begin{figure}[b!]
    \centering
    \includegraphics[width=1\linewidth]{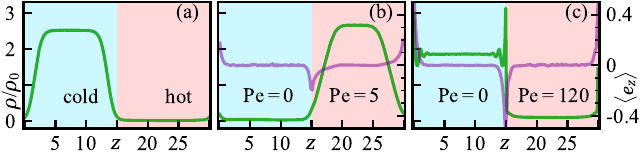}
    \caption{Steady-state density profiles ($\rho/\rho_0$, green, left axis) and $\langle e_z\rangle$, purple, right axis) for: (a) A passive condensate in a thermal gradient. (b) An attractive active condensate in an activity gradient. (c) A MIPS cluster in an activity gradient. Simulation animations are presented in Movie~2 of SM.
    \chg{In the passive reference simulation, the temperature gradient is generated by adding kinetic energy to the hot slab and removing the same amount from the cold slab, producing a stationary temperature profile along \(z\) (see Sec.~SV of the SM for details).}}
    \label{fig:p}
\end{figure}

Is interfacial polarization the dominant mechanism driving this behavior? To address this question, we compare an active condensate in an activity gradient to a passive condensate in a thermal gradient (computational details are provided in SM, Sec.~SV).
In active systems, high-activity regions are analogous to hot regions in passive systems~\cite{takatori2015, ginot2015, speck2016}. Models based on interfacial pressure predict that both droplets should migrate toward the colder or low-activity region~\cite{takatori2015, ginot2015, speck2016, dai2020, cunha2024}.
Our simulations recapitulate this expected behavior for the passive droplet (see Fig.~\ref{fig:p}(a)). In contrast, the active condensate migrates toward the high-activity region (Fig.~\ref{fig:p}(b)).
\chg{This opposite migration direction shows that the behavior cannot be captured by a passive thermophoretic droplet picture or by a scalar effective-temperature mapping.
The essential difference is that attractive active condensates develop a distinct interfacial orientational polarization, which produces an active body-force density \(b_{\rm act}(z)=f_{\rm a}(z)\rho(z)\langle e_z(z)\rangle\).
The resulting active pressure imbalance can drive motion toward higher activity, opposite to the passive case (see Fig.~S7 of SM for pressure profiles).}
We also perform simulations for ABP clusters formed by motility-induced phase separation (MIPS) using hard-sphere ABPs (without pair-attractions).
For these clusters, the propulsion direction at the liquid-gas interface points toward the condensed phase~\cite{omar2023, li2025surface}, and consequently, the clusters migrate toward regions of low activity (Fig.~\ref{fig:p}(c)).
These results reveal that the orientational ordering of ABPs at the interface is the dominant driving force.
In panel (c), the interface exhibits strong oscillations in the density profile. Particles that move from the region with ${\rm Pe}=120$ into regions with ${\rm Pe}=0$ lose their self-propulsion and pile up near the interface, creating an over-compressed first layer. Steric packing behind this layer then produces a sequence of oscillations in the density profile. \chg{The density profile near the interface is reminiscent of packing-induced layering in dense passive colloidal systems confined by soft repulsive walls; see Fig.~S8 in the SM and Refs.~\cite{Tschopp2025, Tschopp2026}.}



\begin{figure}[t!]
    \centering
    \includegraphics[width=1\linewidth]{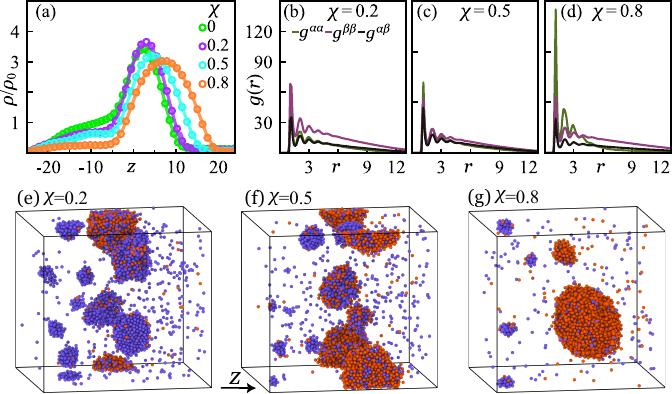}
    \caption{Binary mixtures of passive particles and \chg{ABPs}. A fraction $1-\chi$ of ABPs are $\alpha$-type with $\epsilon^{\alpha\alpha}=4$ and a fraction $\chi$ are passive $\beta$-type with $\epsilon^{\beta\beta}=2$. (a) Steady–state density $\rho(z)$. (b–d) $g^{ij}(r)$ with (e-g) corresponding snapshots, respectively. ABPs are in blue and passive particles in orange.
    }
    \label{fig:5}
\end{figure}

Biological condensates are often multi-component, including active and passive components~\cite{heisenberg2013, ladoux2017, dance2021}. 
\chg{As an exploratory extension, we examine whether activity-gradient-induced localization persists in binary mixtures of active and passive particles.}
To this end, we perform simulations of a binary mixture containing $N=5000$ particles of two types, $\alpha$ and $\beta$, where $\chi$ represents the fraction of $\beta$ particles. Type-$\alpha$ particles are active (ABPs) with a self-interaction strength of $\epsilon^{\alpha \alpha}=4$, and type-$\beta$ particles are passive with $\epsilon^{\beta \beta}=2$ (Fig.~\ref{fig:5}).
The cross-interaction strength between the two types is set to $\epsilon^{\alpha\beta}=3$, with all interactions governed by the WF potential.
The activity field is given by $f_{\rm a}(z) = 20(z + 25)/50$, and the system is non-periodic along the $z$-axis.
Figure~\ref{fig:5}(a) shows the steady-state particle density profiles along the $z$-axis for varying $\chi$.
The case of $\chi=0$ (only $\alpha$-particles) reproduces the results shown in Fig.~\ref{fig:4} for \chg{a single-component active system}.
As $\chi$ increases, the peak of the total particle density shifts monotonically toward the higher activity region (i.e., higher $z$), since passive particles suppress the evaporation of ABPs. Thus, larger clusters are stabilized, which can migrate further into high-activity regions.

\chg{The pair correlation functions \(g(r)\) and simulation snapshots [Figs.~\ref{fig:5}(b)--(g)] show that the two components are not homogeneously mixed.} Visually, the system forms core--shell-like structures at $\chi=0.5$ and $0.8$. The strongly attractive ABPs (blue) condense into dense cores that are enveloped by a periphery enriched with weakly attractive passive particles (orange).
\chg{The large peak in \(g^{\alpha\alpha}(r)\), together with the smaller \(g^{\alpha\beta}(r)\) peak, indicates enhanced correlations among the strongly attractive ABPs.}
Moreover, as $\chi$  increases, the $g^{\alpha\alpha}$ peak grows, suggesting that the \(\alpha\)-rich domains become more densely packed.
\chg{The active--passive mixture discussed here can be viewed as the limiting case in which the weakly cohesive \(\beta\) component has zero activity.
As a related complementary case, we also consider mixtures in which both components are active but differ in cohesive strength (cf. Fig.~S9 of SM). These simulations show a qualitatively similar enrichment of the more cohesive component in the dense interior.}
\chg{Phenomenologically, this structural arrangement shares a visual resemblance to the differential adhesion hypothesis fundamental to cell sorting and tissue morphogenesis~\cite{heisenberg2013, ladoux2017, dance2021, foty2005, chanson2011, cerchiari2015, toda2018, wang2021}, which has likewise inspired robotic swarm control~\cite{santos2020, ceron2023, pan2024}.}

\section{Conclusions}
\chg{Cohesive active assemblies can display emergent, group-level responses to spatial cues.
For example, in biological active matter, groups of cells can respond collectively to spatial guidance cues that are weak or absent at the single-cell level~\cite{ellison2016, giampieri2009, theveneau2010, malet2015, bussmann2015}.
Our results show that group-level responses to spatial gradients can arise in a minimal active-matter setting, where cohesion and spatially varying activity generate condensate migration, localization, turnover, and cyclic repositioning.}
Interparticle attraction drives liquid-gas transition, and the competition between self–propulsion and attraction determines their directional transport.
We find that the direction of motion is determined by the preferential orientation of ABPs at the droplet interface, while the interior remains disordered.
In the strong-attraction regime, stable, giant clusters form and are driven to regions of highest activity.
At intermediate attraction strengths, a dynamic living state emerges, characterized by clusters that perpetually fragment and reassemble. These living clusters reside in intermediate activity regions.
Adding passive, less cohesive particles stabilizes the living clusters, which migrate toward higher-activity regions.

Our findings contrast with MIPS clusters, where inward-pointing particles at the interface drive condensates toward low activity. This highlights that a new theory, beyond MIPS-based descriptions, is required to capture the phase transitions of attractive ABPs in inhomogeneous systems.
Furthermore, previous studies of two-dimensional systems have reported flocking for attractive ABPs~\cite{caprini2023}, though finite-size analysis indicates that this does not persist as $N\!\to\!\infty$~\cite{mahault2023}.
Our three-dimensional simulations show no evidence of global flocking, even when activity gradients drive directed condensate migration.
\chg{Although we focus here on linear activity gradients, the pressure-balance mechanism does not rely on the gradient being strictly linear. For a smooth monotonic activity profile, we expect condensates to localize near positions where the activity-induced pressure imbalance between the inner and outer interfaces is balanced by cohesive and ideal-pressure contributions. Changing the gradient shape or length scale should therefore shift the balance point and modify the migration rate, while non-monotonic activity profiles may generate multiple localization regions.}
Looking forward, our results could be experimentally tested by engineering synthetic active condensates based on platforms like enzyme-powered nanomotors~\cite{song2021, mason2017}, light-activated Janus particles~\cite{palacci2013}, and robotic swarms~\cite{santos2020, ceron2023, pan2024}.

\section*{Acknowledgments}
This work was supported by Deutsche Forschungsgemeinschaft (DFG) under Project No. 561963765 (H.V.) and No. 525864799 (A.S.). J.U.S.~thanks the cluster of excellence “Physics of Life” at TU Dresden for its support. We gratefully acknowledge the NHR Center at TU Dresden for providing high-performance computing resources.

\bibliography{references}

@article{bechinger2016,
  title={Active particles in complex and crowded environments},
  author={Bechinger, Clemens and Di Leonardo, Roberto and L{\"o}wen, Hartmut and Reichhardt, Charles and Volpe, Giorgio and Volpe, Giovanni},
  journal={Rev. Mod. Phys.},
  volume={88},
  number={4},
  pages={045006},
  year={2016},
  publisher={APS},
  doi={10.1103/RevModPhys.88.045006}
}

@article{theveneau2010,
  title={Collective chemotaxis requires contact-dependent cell polarity},
  author={Theveneau, Eric and Marchant, Lorena and Kuriyama, Sei and Gull, Mazhar and Moepps, Barbara and Parsons, Maddy and Mayor, Roberto},
  journal={Dev. Cell},
  volume={19},
  number={1},
  pages={39--53},
  year={2010},
  publisher={Elsevier},
  doi={10.1016/j.devcel.2010.06.012}
}

@article{caprini2023,
  title={Flocking without alignment interactions in attractive active brownian particles},
  author={Caprini, Lorenzo and L{\"o}wen, Hartmut},
  journal={Phys. Rev. Lett.},
  volume={130},
  number={14},
  pages={148202},
  year={2023},
  publisher={APS},
  doi={10.1103/PhysRevLett.130.148202}
}

@article{wang2020,
  title={The Lennard-Jones potential: when (not) to use it},
  author={Wang, Xipeng and Ram{\'\i}rez-Hinestrosa, Sim{\'o}n and Dobnikar, Jure and Frenkel, Daan},
  journal={Phys. Chem. Chem. Phys.},
  volume={22},
  number={19},
  pages={10624--10633},
  year={2020},
  publisher={Royal Society of Chemistry},
  doi={10.1039/C9CP05445F}
}

@article{palacci2013,
  title={Living crystals of light-activated colloidal surfers},
  author={Palacci, Jeremie and Sacanna, Stefano and Steinberg, Asher Preska and Pine, David J and Chaikin, Paul M},
  journal={Science},
  volume={339},
  number={6122},
  pages={936--940},
  year={2013},
  publisher={American Association for the Advancement of Science},
  doi={10.1126/science.1230020}
}

@article{ellison2016,
  title={Cell--cell communication enhances the capacity of cell ensembles to sense shallow gradients during morphogenesis},
  author={Ellison, David and Mugler, Andrew and Brennan, Matthew D and Lee, Sung Hoon and Huebner, Robert J and Shamir, Eliah R and Woo, Laura A and Kim, Joseph and Amar, Patrick and Nemenman, Ilya and others},
  journal={Proc. Natl. Acad. Sci.},
  volume={113},
  number={6},
  pages={E679--E688},
  year={2016},
  publisher={National Acad Sciences},
  doi={10.1073/pnas.1516503113}
}

@article{malet2015,
  title={Collective cell motility promotes chemotactic prowess and resistance to chemorepulsion},
  author={Malet-Engra, Gema and Yu, Weimiao and Oldani, Amanda and Rey-Barroso, Javier and Gov, Nir S and Scita, Giorgio and Dupr{\'e}, Lo{\"\i}c},
  journal={Curr. Biol.},
  volume={25},
  number={2},
  pages={242--250},
  year={2015},
  publisher={Elsevier},
  doi={10.1016/j.cub.2014.11.030}
}

@article{mahault2023,
  title={Comment on" Flocking without Alignment Interactions in Attractive Active Brownian Particles [arXiv: 2303.07746]"},
  author={Mahault, Beno{\^\i}t},
  journal={arXiv preprint arXiv:2309.00015},
  year={2023},
  doi={10.48550/arXiv.2309.00015}
}

@article{schwarz2012,
  title={Phase separation and rotor self-assembly in active particle suspensions},
  author={Schwarz-Linek, J and Valeriani, C and Cacciuto, A and Cates, ME and Marenduzzo, D and Morozov, AN and Poon, WCK},
  journal={Proc. Natl. Acad. Sci.},
  volume={109},
  number={11},
  pages={4052--4057},
  year={2012},
  publisher={National Academy of Sciences},
  doi={10.1073/pnas.1116334109}
}

@article{foty2005,
  title={The differential adhesion hypothesis: a direct evaluation},
  author={Foty, Ramsey A and Steinberg, Malcolm S},
  journal={Dev. Biol.},
  volume={278},
  number={1},
  pages={255--263},
  year={2005},
  publisher={Elsevier},
  doi={10.1016/j.ydbio.2004.11.012}
}

@article{hyman2014,
  title={Liquid-liquid phase separation in biology},
  author={Hyman, Anthony A and Weber, Christoph A and J{\"u}licher, Frank},
  journal={Annu. Rev. Cell Dev. Biol.},
  volume={30},
  number={1},
  pages={39--58},
  year={2014},
  publisher={Annual Reviews},
  doi={10.1146/annurev-cellbio-100913-013325}
}

@article{banani2017,
  title={Biomolecular condensates: organizers of cellular biochemistry},
  author={Banani, Salman F and Lee, Hyun O and Hyman, Anthony A and Rosen, Michael K},
  journal={Nat. Rev. Mol. Cell Biol.},
  volume={18},
  number={5},
  pages={285--298},
  year={2017},
  publisher={Nature Publishing Group UK London},
  doi={10.1038/nrm.2017.7}
}

@article{weber2019,
  title={Physics of active emulsions},
  author={Weber, Christoph A and Zwicker, David and J{\"u}licher, Frank and Lee, Chiu Fan},
  journal={Rep. Prog. Phys.},
  volume={82},
  number={6},
  pages={064601},
  year={2019},
  publisher={IOP Publishing},
  doi={10.1088/1361-6633/ab052b}
}

@article{giampieri2009,
  title={Localized and reversible TGF$\beta$ signalling switches breast cancer cells from cohesive to single cell motility},
  author={Giampieri, Silvia and Manning, Cerys and Hooper, Steven and Jones, Louise and Hill, Caroline S and Sahai, Erik},
  journal={Nat. Cell Biol.},
  volume={11},
  number={11},
  pages={1287--1296},
  year={2009},
  publisher={Nature Publishing Group UK London},
  doi={10.1038/ncb1973}
}

@article{bussmann2015,
  title={Chemokine-guided cell migration and motility in zebrafish development},
  author={Bussmann, Jeroen and Raz, Erez},
  journal={EMBO J.},
  volume={34},
  number={10},
  pages={1309--1318},
  year={2015},
  doi={10.15252/embj.201490105}
}

@article{scarpa2016,
  title={Collective cell migration in development},
  author={Scarpa, Elena and Mayor, Roberto},
  journal={J. Cell Biol.},
  volume={212},
  number={2},
  pages={143--155},
  year={2016},
  publisher={The Rockefeller University Press},
  doi={10.1083/jcb.201508047 }
}

@article{ewald2008,
  title={Collective epithelial migration and cell rearrangements drive mammary branching morphogenesis},
  author={Ewald, Andrew J and Brenot, Audrey and Duong, Myhanh and Chan, Bianca S and Werb, Zena},
  journal={Dev. Cell.},
  volume={14},
  number={4},
  pages={570--581},
  year={2008},
  publisher={Elsevier},
  doi={10.1016/j.devcel.2008.03.003 }
}

@article{li2013,
  title={Collective cell migration: Implications for wound healing and cancer invasion},
  author={Li, Li and He, Yong and Zhao, Min and Jiang, Jianxin},
  journal={Burns Trauma},
  volume={1},
  number={1},
  pages={2321--3868},
  year={2013},
  publisher={Oxford University Press},
  doi={10.4103/2321-3868.113331}
}

@article{riahi2012,
  title={Advances in wound-healing assays for probing collective cell migration},
  author={Riahi, Reza and Yang, Yongliang and Zhang, Donna D and Wong, Pak Kin},
  journal={J. Lab. Autom.},
  volume={17},
  number={1},
  pages={59--65},
  year={2012},
  publisher={SAGE Publications Sage CA: Los Angeles, CA},
 doi={10.1177/2211068211426550}
}

@article{prymidis2016,
  title={Vapour-liquid coexistence of an active Lennard-Jones fluid},
  author={Prymidis, Vasileios and Paliwal, Siddharth and Dijkstra, Marjolein and Filion, Laura},
  journal={J. Chem. Phys.},
  volume={145},
  number={12},
  year={2016},
  publisher={AIP Publishing},
  doi={10.1063/1.4963191}
}

@article{paliwal2017,
  title={Non-equilibrium surface tension of the vapour-liquid interface of active Lennard-Jones particles},
  author={Paliwal, Siddharth and Prymidis, Vasileios and Filion, Laura and Dijkstra, Marjolein},
  journal={J. Chem. Phys.},
  volume={147},
  number={8},
  year={2017},
  publisher={AIP Publishing},
  doi={10.1063/1.4989764}
}

@article{omar2020,
  title={Microscopic origins of the swim pressure and the anomalous surface tension of active matter},
  author={Omar, Ahmad K and Wang, Zhen-Gang and Brady, John F},
  journal={Phys. Rev. E},
  volume={101},
  number={1},
  pages={012604},
  year={2020},
  publisher={APS},
  doi={10.1103/PhysRevE.101.012604}
}

@article{vuijk2022,
  title={Active colloidal molecules in activity gradients},
  author={Vuijk, Hidde D and Klempahn, Sophie and Merlitz, Holger and Sommer, Jens-Uwe and Sharma, Abhinav},
  journal={Phys. Rev. E},
  volume={106},
  number={1},
  pages={014617},
  year={2022},
  publisher={APS},
  doi={10.1103/PhysRevE.106.014617}
}

@article{ginot2018,
  title={Aggregation-fragmentation and individual dynamics of active clusters},
  author={Ginot, F{\'e}lix and Theurkauff, Isaac and Detcheverry, F and Ybert, Christophe and Cottin-Bizonne, C{\'e}cile},
  journal={Nat. Commun.},
  volume={9},
  number={1},
  pages={696},
  year={2018},
  publisher={Nature Publishing Group UK London},
  doi={10.1038/s41467-017-02625-7}
}

@article{mognetti2013,
  title={Living clusters and crystals from low-density suspensions of active colloids},
  author={Mognetti, Bortolo Matteo and {\v{S}}ari{\'c}, An{\dj}ela and Angioletti-Uberti, Stefano and Cacciuto, A and Valeriani, C and Frenkel, Daan},
  journal={Phys. Rev. Lett.},
  volume={111},
  number={24},
  pages={245702},
  year={2013},
  publisher={APS},
  doi={10.1103/PhysRevLett.111.245702}
}

@article{caprini2024,
  title={Dynamical clustering and wetting phenomena in inertial active matter},
  author={Caprini, Lorenzo and Breoni, Davide and Ldov, Anton and Scholz, Christian and L{\"o}wen, Hartmut},
  journal={Commun. Phys.},
  volume={7},
  number={1},
  pages={343},
  year={2024},
  publisher={Nature Publishing Group UK London},
  doi={10.1038/s42005-024-01835-y}
}

@article{alert2020,
  title={Physical models of collective cell migration},
  author={Alert, Ricard and Trepat, Xavier},
  journal={Annu. Rev. Condens. Matter Phys.},
  volume={11},
  number={1},
  pages={77--101},
  year={2020},
  publisher={Annual Reviews},
  doi={10.1146/annurev-conmatphys-031218-013516}
}

@article{stukowski2009,
  title={Visualization and analysis of atomistic simulation data with OVITO--the Open Visualization Tool},
  author={Stukowski, Alexander},
  journal={Model. Simul. Mater. Sci. Eng.},
  volume={18},
  number={1},
  pages={015012},
  year={2009},
  publisher={IOP Publishing},
  doi={10.1088/0965-0393/18/1/015012}
}

@Article{LAMMPS,
  author = "A. P. Thompson and H. M. Aktulga and R. Berger and 
     D. S. Bolintineanu and W. M. Brown and P. S. Crozier and
     P. J. in 't Veld and A. Kohlmeyer and S. G. Moore and T. D. Nguyen and
     R. Shan and M. J. Stevens and J. Tranchida and C. Trott and S. J. Plimpton",
  title = "{LAMMPS} - a flexible simulation tool for
     particle-based materials modeling at the 
     atomic, meso, and continuum scales",
  journal = "Comp. Phys. Comm.",
  volume =  "271",
  pages =   "108171",
  year =    "2022",
  doi = "10.1016/j.cpc.2021.108171"
}

@article{omori2024,
  title={Molecular anatomy of the pressure anisotropy in the interface of one and two component fluids: Local thermodynamic description of the interfacial tension},
  author={Omori, Takeshi and Yamaguchi, Yasutaka},
  journal={J. Chem. Phys.},
  volume={161},
  number={20},
  year={2024},
  publisher={AIP Publishing},
  doi={10.1063/5.0235858}
}

@article{ziepke2022,
  title={Multi-scale organization in communicating active matter},
  author={Ziepke, Alexander and Maryshev, Ivan and Aranson, Igor S and Frey, Erwin},
  journal={Nat. Commun.},
  volume={13},
  number={1},
  pages={6727},
  year={2022},
  publisher={Nature Publishing Group UK London},
  doi={doi.org/10.1038/s41467-022-34484-2}
}

@article{omar2023,
  title={Mechanical theory of nonequilibrium coexistence and motility-induced phase separation},
  author={Omar, Ahmad K and Row, Hyeongjoo and Mallory, Stewart A and Brady, John F},
  journal={Proc. Natl. Acad. Sci.},
  volume={120},
  number={18},
  pages={e2219900120},
  year={2023},
  publisher={National Academy of Sciences},
  doi={10.1073/pnas.2219900120}
}

@article{dai2020,
  title={Migration of liquid bridges at the interface of spheres and plates with an imposed thermal gradient},
  author={Dai, Qingwen and Chong, Zhejun and Huang, Wei and Wang, Xiaolei},
  journal={Langmuir},
  volume={36},
  number={22},
  pages={6268--6276},
  year={2020},
  publisher={ACS Publications},
  doi={10.1021/acs.langmuir.9b03951}
}

@article{jambon2024,
  title={Phase-separated droplets swim to their dissolution},
  author={Jambon-Puillet, Etienne and Testa, Andrea and Lorenz, Charlotta and Style, Robert W and Rebane, Aleksander A and Dufresne, Eric R},
  journal={Nat. Commun.},
  volume={15},
  number={1},
  pages={3919},
  year={2024},
  publisher={Nature Publishing Group UK London},
  doi={10.1038/s41467-024-47889-y}
}

@article{cunha2024,
  title={Influence of liquid--vapor phase change on the self-propelled motion of droplets on wettability gradient surfaces},
  author={Cunha, Vitor HC and Dorao, Carlos A and Fernandino, Maria},
  journal={Phys. Fluids},
  volume={36},
  number={12},
  year={2024},
  publisher={AIP Publishing},
  doi={10.1063/5.0239562}
}

@article{takatori2015,
  title={A theory for the phase behavior of mixtures of active particles},
  author={Takatori, Sho C and Brady, John F},
  journal={Soft Matter},
  volume={11},
  number={40},
  pages={7920--7931},
  year={2015},
  publisher={Royal Society of Chemistry},
  doi={10.1039/C5SM01792K}
}

@article{ginot2015,
  title={Nonequilibrium equation of state in suspensions of active colloids},
  author={Ginot, F{\'e}lix and Theurkauff, Isaac and Levis, Demian and Ybert, Christophe and Bocquet, Lyd{\'e}ric and Berthier, Ludovic and Cottin-Bizonne, C{\'e}cile},
  journal={Phys. Rev. X},
  volume={5},
  number={1},
  pages={011004},
  year={2015},
  publisher={APS},
  doi={10.1103/PhysRevX.5.011004}
}

@article{speck2016,
  title={Stochastic thermodynamics for active matter},
  author={Speck, Thomas},
  journal={Europhys. Lett. },
  volume={114},
  number={3},
  pages={30006},
  year={2016},
  publisher={IOP Publishing},
  doi={10.1209/0295-5075/114/30006}
}

@article{saha2016,
  title={Polar positioning of phase-separated liquid compartments in cells regulated by an mRNA competition mechanism},
  author={Saha, Shambaditya and Weber, Christoph A and Nousch, Marco and Adame-Arana, Omar and Hoege, Carsten and Hein, Marco Y and Osborne-Nishimura, Erin and Mahamid, Julia and Jahnel, Marcus and Jawerth, Louise and others},
  journal={Cell},
  volume={166},
  number={6},
  pages={1572--1584},
  year={2016},
  publisher={Elsevier},
  doi={10.1016/j.cell.2016.08.006 }
}

@article{snead2019,
  title={The control centers of biomolecular phase separation: how membrane surfaces, PTMs, and active processes regulate condensation},
  author={Snead, Wilton T and Gladfelter, Amy S},
  journal={Mol. Cell.},
  volume={76},
  number={2},
  pages={295--305},
  year={2019},
  publisher={Elsevier},
  doi={10.1016/j.molcel.2019.09.016}
}

@article{brangwynne2011,
  title={Active liquid-like behavior of nucleoli determines their size and shape in Xenopus laevis oocytes},
  author={Brangwynne, Clifford P and Mitchison, Timothy J and Hyman, Anthony A},
  journal={Proc. Natl. Acad. Sci.},
  volume={108},
  number={11},
  pages={4334--4339},
  year={2011},
  publisher={National Academy of Sciences},
  doi={10.1073/pnas.1017150108}
}

@article{feric2016,
  title={Coexisting liquid phases underlie nucleolar subcompartments},
  author={Feric, Marina and Vaidya, Nilesh and Harmon, Tyler S and Mitrea, Diana M and Zhu, Lian and Richardson, Tiffany M and Kriwacki, Richard W and Pappu, Rohit V and Brangwynne, Clifford P},
  journal={Cell},
  volume={165},
  number={7},
  pages={1686--1697},
  year={2016},
  publisher={Elsevier},
  doi={10.1016/j.cell.2016.04.047 }
}

@article{jain2016,
  title={ATPase-modulated stress granules contain a diverse proteome and substructure},
  author={Jain, Saumya and Wheeler, Joshua R and Walters, Robert W and Agrawal, Anurag and Barsic, Anthony and Parker, Roy},
  journal={Cell},
  volume={164},
  number={3},
  pages={487--498},
  year={2016},
  publisher={Elsevier},
  doi={10.1016/j.cell.2015.12.038}
}

@article{linsenmeier2022,
  title={Dynamic arrest and aging of biomolecular condensates are modulated by low-complexity domains, RNA and biochemical activity},
  author={Linsenmeier, Miriam and Hondele, Maria and Grigolato, Fulvio and Secchi, Eleonora and Weis, Karsten and Arosio, Paolo},
  journal={Nat. Commun.},
  volume={13},
  number={1},
  pages={3030},
  year={2022},
  publisher={Nature Publishing Group UK London},
  doi={10.1038/s41467-022-30521-2}
}

@article{song2021,
  title={Engineering transient dynamics of artificial cells by stochastic distribution of enzymes},
  author={Song, Shidong and Mason, Alexander F and Post, Richard AJ and De Corato, Marco and Mestre, Rafael and Yewdall, N Amy and Cao, Shoupeng and van der Hofstad, Remco W and Sanchez, Samuel and Abdelmohsen, Loai KEA and others},
  journal={Nat. Commun.},
  volume={12},
  number={1},
  pages={6897},
  year={2021},
  publisher={Nature Publishing Group UK London},
  doi={10.1038/s41467-021-27229-0}
}

@article{mason2017,
  title={Hierarchical self-assembly of a copolymer-stabilized coacervate protocell},
  author={Mason, Alexander F and Buddingh’, Bastiaan C and Williams, David S and Van Hest, Jan CM},
  journal={J. Am. Chem. Soc.},
  volume={139},
  number={48},
  pages={17309--17312},
  year={2017},
  publisher={ACS Publications},
  doi={10.1021/jacs.7b10846}
}

@article{schwarz2010,
  title={Polymer-induced phase separation in Escherichia coli suspensions},
  author={Schwarz-Linek, Jana and Winkler, Alexander and Wilson, Laurence G and Pham, Nhan T and Schilling, Tanja and Poon, Wilson CK},
  journal={Soft Matter},
  volume={6},
  number={18},
  pages={4540--4549},
  year={2010},
  publisher={Royal Society of Chemistry},
  doi={10.1039/C0SM00214C}
}

@article{mayor2010,
  title={Keeping in touch with contact inhibition of locomotion},
  author={Mayor, Roberto and Carmona-Fontaine, Carlos},
  journal={Trends Cell Biol.},
  volume={20},
  number={6},
  pages={319--328},
  year={2010},
  publisher={Elsevier},
  doi={10.1016/j.tcb.2010.03.005 }
}

@article{guilhas2020,
  title={ATP-driven separation of liquid phase condensates in bacteria},
  author={Guilhas, Baptiste and Walter, Jean-Charles and Rech, Jerome and David, Gabriel and Walliser, Nils Ole and Palmeri, John and Mathieu-Demaziere, Celine and Parmeggiani, Andrea and Bouet, Jean-Yves and Le Gall, Antoine and others},
  journal={Mol. Cell.},
  volume={79},
  number={2},
  pages={293--303},
  year={2020},
  publisher={Elsevier},
  doi={10.1016/j.molcel.2020.06.034 }
}

@article{lin2015,
  title={Interplay between chemotaxis and contact inhibition of locomotion determines exploratory cell migration},
  author={Lin, Benjamin and Yin, Taofei and Wu, Yi I and Inoue, Takanari and Levchenko, Andre},
  journal={Nat. Commun.},
  volume={6},
  number={1},
  pages={6619},
  year={2015},
  publisher={Nature Publishing Group UK London},
  doi={10.1038/ncomms7619}
}

@article{miskolci2021,
  title={Cell migration guided by cell--cell contacts in innate immunity},
  author={Miskolci, Veronika and Klemm, Lucas C and Huttenlocher, Anna},
  journal={Trends Cell Biol.},
  volume={31},
  number={2},
  pages={86--94},
  year={2021},
  publisher={Elsevier},
  doi={10.1016/j.tcb.2020.11.002}
}

@article{carmona2008,
  title={Contact inhibition of locomotion in vivo controls neural crest directional migration},
  author={Carmona-Fontaine, Carlos and Matthews, Helen K and Kuriyama, Sei and Moreno, Mauricio and Dunn, Graham A and Parsons, Maddy and Stern, Claudio D and Mayor, Roberto},
  journal={Nature},
  volume={456},
  number={7224},
  pages={957--961},
  year={2008},
  publisher={Nature Publishing Group UK London},
  doi={10.1038/nature07441}
}

@article{li2019,
  title={In vivo quantitative imaging provides insights into trunk neural crest migration},
  author={Li, Yuwei and Vieceli, Felipe M and Gonzalez, Walter G and Li, Ang and Tang, Weiyi and Lois, Carlos and Bronner, Marianne E},
  journal={Cell Rep.},
  volume={26},
  number={6},
  pages={1489--1500},
  year={2019},
  publisher={Elsevier},
  doi={10.1016/j.celrep.2019.01.039}
}

@article{cerchiari2015,
  title={A strategy for tissue self-organization that is robust to cellular heterogeneity and plasticity},
  author={Cerchiari, Alec E and Garbe, James C and Jee, Noel Y and Todhunter, Michael E and Broaders, Kyle E and Peehl, Donna M and Desai, Tejal A and LaBarge, Mark A and Thomson, Matthew and Gartner, Zev J},
  journal={Proc. Natl. Acad. Sci.},
  volume={112},
  number={7},
  pages={2287--2292},
  year={2015},
  publisher={National Academy of Sciences},
  doi={10.1073/pnas.1410776112}
}

@article{wang2021,
  title={Budding epithelial morphogenesis driven by cell-matrix versus cell-cell adhesion},
  author={Wang, Shaohe and Matsumoto, Kazue and Lish, Samantha R and Cartagena-Rivera, Alexander X and Yamada, Kenneth M},
  journal={Cell},
  volume={184},
  number={14},
  pages={3702--3716},
  year={2021},
  publisher={Elsevier},
  doi={10.1016/j.cell.2021.05.015 }
}

@article{toda2018,
  title={Programming self-organizing multicellular structures with synthetic cell-cell signaling},
  author={Toda, Satoshi and Blauch, Lucas R and Tang, Sindy KY and Morsut, Leonardo and Lim, Wendell A},
  journal={Science},
  volume={361},
  number={6398},
  pages={156--162},
  year={2018},
  publisher={American Association for the Advancement of Science},
  doi={10.1126/science.aat0271}
}

@article{chanson2011,
  title={Self-organization is a dynamic and lineage-intrinsic property of mammary epithelial cells},
  author={Chanson, Lea and Brownfield, Douglas and Garbe, James C and Kuhn, Irene and Stampfer, Martha R and Bissell, Mina J and LaBarge, Mark A},
  journal={Proc. Natl. Acad. Sci.},
  volume={108},
  number={8},
  pages={3264--3269},
  year={2011},
  publisher={National Academy of Sciences},
  doi={10.1073/pnas.1019556108}
}

@article{heisenberg2013,
  title={Forces in tissue morphogenesis and patterning},
  author={Heisenberg, Carl-Philipp and Bella{\"\i}che, Yohanns},
  journal={Cell},
  volume={153},
  number={5},
  pages={948--962},
  year={2013},
  publisher={Elsevier},
  doi={10.1016/j.cell.2013.05.008}
}

@article{ladoux2017,
  title={Mechanobiology of collective cell behaviours},
  author={Ladoux, Benoit and M{\`e}ge, Ren{\'e}-Marc},
  journal={Nat. Rev. Mol. Cell Biol.},
  volume={18},
  number={12},
  pages={743--757},
  year={2017},
  publisher={Nature Publishing Group UK London}
}

@article{dance2021,
  title={The secret forces that squeeze and pull life into shape},
  author={Dance, Amber},
  journal={Nature},
  volume={589},
  number={7841},
  pages={186--189},
  year={2021},
  publisher={Nature Publishing Group},
  doi={10.1038/d41586-021-00018-x}
}

@article{ceron2023,
  title={Programmable self-organization of heterogeneous microrobot collectives},
  author={Ceron, Steven and Gardi, Gaurav and Petersen, Kirstin and Sitti, Metin},
  journal={Proc. Natl. Acad. Sci.},
  volume={120},
  number={24},
  pages={e2221913120},
  year={2023},
  publisher={National Academy of Sciences},
  doi={10.1073/pnas.2221913120}
}

@article{pan2024,
  title={Applying the intrinsic principle of cell collectives to program robot swarms},
  author={Pan, Mengyun and Yang, Yongliang and Qin, Xiaoyang and Li, Guangyong and Xi, Ning and Long, Min and Jiang, Lei and Zhao, Tianming and Liu, Lianqing},
  journal={Cell Rep. Phys. Sci.},
  volume={5},
  number={8},
  year={2024},
  publisher={Elsevier},
 doi={10.1016/j.xcrp.2024.102122}
}

@article{santos2020,
  title={Spatial segregative behaviors in robotic swarms using differential potentials},
  author={Santos, Vinicius G and Pires, Anderson G and Alitappeh, Reza J and Rezeck, Paulo AF and Pimenta, Luciano CA and Macharet, Douglas G and Chaimowicz, Luiz},
  journal={Swarm Intell.},
  volume={14},
  number={4},
  pages={259--284},
  year={2020},
  publisher={Springer},
  doi={10.1007/s11721-020-00184-0}
}

@article{li2025surface,
author = {Longfei Li  and Zihao Sun  and Mingcheng Yang },
title = {Surface tension between coexisting phases of active Brownian particles},
journal = {Proc. Natl. Acad. Sci.},
volume = {122},
number = {29},
pages = {e2505010122},
year = {2025},
doi = {10.1073/pnas.2505010122},}

@Article{Tschopp2025,
author ="Tschopp, S. M. and Vahid, H. and Sharma, A. and Brader, J. M.",
title  ="Combining integral equation closures with force density functional theory for the study of inhomogeneous fluids",
journal  ="Soft Matter",
year  ="2025",
volume  ="21",
issue  ="14",
pages  ="2633-2645",
publisher  ="The Royal Society of Chemistry",
doi  ="10.1039/D4SM01262C",
url  ="http://dx.doi.org/10.1039/D4SM01262C",}

@article{Tschopp2026,
author = {Tschopp, S. M. and Vahid, H. and Brader, J. M.},
title = {Routes to the Density Profile and Structural Inconsistency},
journal = {J. Phys. Chem. B},
volume = {130},
number = {4},
pages = {1424-1436},
year = {2026},
doi = {10.1021/acs.jpcb.5c07785},
    note ={PMID: 41556538},
}

@article{patel2017,
  title={ATP as a biological hydrotrope},
  author={Patel, Avinash and Malinovska, Liliana and Saha, Shambaditya and Wang, Jie and Alberti, Simon and Krishnan, Yamuna and Hyman, Anthony A},
  journal={Science},
  volume={356},
  number={6339},
  pages={753--756},
  year={2017},
  publisher={American Association for the Advancement of Science},
  doi={10.1126/science.aaf6846}
}

@article{lafontaine2021,
  title={The nucleolus as a multiphase liquid condensate},
  author={Lafontaine, Denis LJ and Riback, Joshua A and Bascetin, R{\"u}meyza and Brangwynne, Clifford P},
  journal={Nat. Rev. Mol. Cell Biol.},
  volume={22},
  number={3},
  pages={165--182},
  year={2021},
  publisher={Nature Publishing Group UK London},
  doi={https://doi.org/10.1038/s41580-020-0272-6}
}

@article{auschra2021,
  title={Polarization-density patterns of active particles in motility gradients},
  author={Auschra, Sven and Holubec, Viktor and S{\"o}ker, Nicola Andreas and Cichos, Frank and Kroy, Klaus},
  journal={Phys. Rev. E},
  volume={103},
  number={6},
  pages={062601},
  year={2021},
  publisher={APS},
  doi={doi.org/10.1103/PhysRevE.103.062601}
}

@article{soker2021,
  title = {How Activity Landscapes Polarize Microswimmers without Alignment Forces},
  author = {S\"oker, Nicola Andreas and Auschra, Sven and Holubec, Viktor and Kroy, Klaus and Cichos, Frank},
  journal = {Phys. Rev. Lett.},
  volume = {126},
  issue = {22},
  pages = {228001},
  numpages = {6},
  year = {2021},
  month = {6},
  publisher = {American Physical Society},
  doi = {10.1103/PhysRevLett.126.228001},
  url = {https://link.aps.org/doi/10.1103/PhysRevLett.126.228001}
}

@article{row2020,
  title = {Reverse osmotic effect in active matter},
  author = {Row, Hyeongjoo and Brady, John F.},
  journal = {Phys. Rev. E},
  volume = {101},
  issue = {6},
  pages = {062604},
  numpages = {9},
  year = {2020},
  month = {6},
  publisher = {American Physical Society},
  doi = {10.1103/PhysRevE.101.062604},
  url = {https://link.aps.org/doi/10.1103/PhysRevE.101.062604}
}

@article{Ikeshoji01012003,
author = {Tamio Ikeshoji and BjØrn Hafskjold and Hilde Furuholt},
title = {Molecular-level Calculation Scheme for Pressure in Inhomogeneous Systems of Flat and Spherical Layers},
journal = {Mol. Simul.},
volume = {29},
number = {2},
pages = {101--109},
year = {2003},
publisher = {Taylor \& Francis},
doi = {10.1080/102866202100002518a},
}

@article{wedlich2018,
    author = {Wedlich-Söldner, Roland and Betz, Timo},
    title = {Self-organization: the fundament of cell biology},
    journal = {Philos. Trans. R. Soc. B},
    volume = {373},
    number = {1747},
    pages = {20170103},
    year = {2018},
    month = {04},
    issn = {0962-8436},
    doi = {10.1098/rstb.2017.0103},
}

@article{karsenti2008,
  title={Self-organization in cell biology: a brief history},
  author={Karsenti, Eric},
  journal={Nat. Rev. Mol. Cell Biol.},
  volume={9},
  number={3},
  pages={255--262},
  year={2008},
  publisher={Nature Publishing Group UK London},
  doi={doi.org/10.1038/nrm2357}
}

@Article{coupe2026,
author ="Coupe, Sebastian and Fakhri, Nikta",
title  ="Nonequilibrium phases of a biomolecular condensate facilitated by enzyme activity",
journal  ="Soft Matter",
year  ="2026",
volume  ="22",
issue  ="14",
pages  ="2656-2665",
publisher  ="The Royal Society of Chemistry",
doi  ="10.1039/D5SM01106J",
url  ="http://dx.doi.org/10.1039/D5SM01106J",}

\foreach \x in {1,...,8}
{%
\clearpage
\includepdf[pages={\x}]{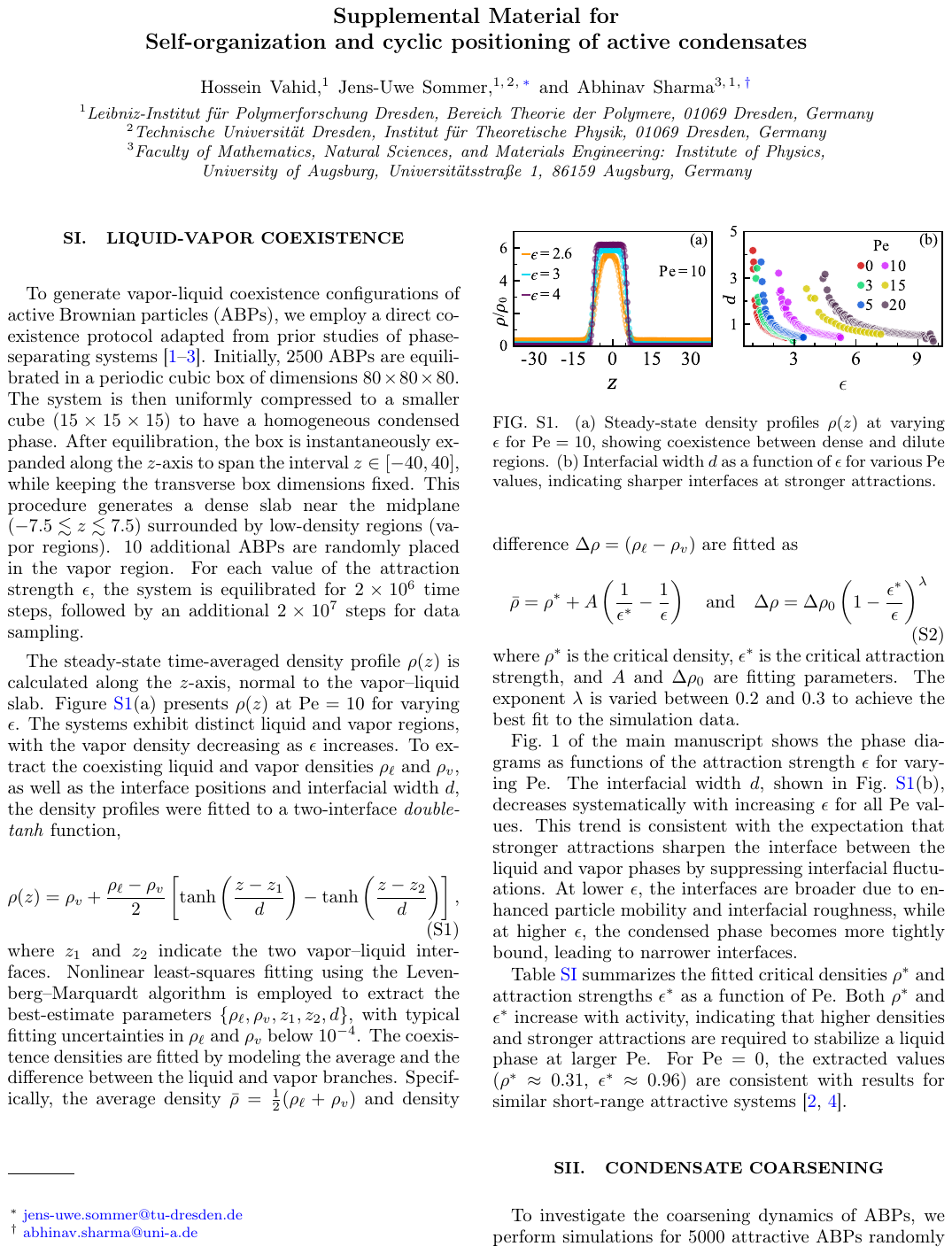} 
}

\end{document}